\documentclass[10pt,a4paper]{article}

\usepackage{indentfirst}                        

\newenvironment{resum}{\begin{quote}\small}{\end{quote}}

\font\mybold=cmmib10
\chardef\Myxi="18
\def\boldxi{\hbox{\mybold\Myxi}}

\font\mybold=cmmib10
\chardef\Mychi="1F
\def\boldchi{\hbox{\mybold\Mychi}}

\newcommand{\bfsf}[1]{\textsf{\textbf{#1}}}

\begin{document}

\thispagestyle{plain}           

\begin{center}


{\LARGE\bfsf{Killing Tensors from Conformal Killing Vectors}}

\bigskip


\textbf{Alan Barnes}$^1$, \textbf{Brian Edgar}$^2$ and
\textbf{Raffaele Rani}$^2$


$^1$\textsl{School of Engineering \& Applied Science,\\
Aston University, Birmingham, B4 7ET, U.K.} \\
$^2$\textsl{Department of Mathematics, University of Link\"oping,\\
SE 581 83 Link\"oping, Sweden}
\end{center}

\medskip

\begin{resum}
Some years ago Koutras presented a method of constructing a conformal Killing
tensor from a pair of orthogonal conformal Killing vectors. When the
vector associated with the conformal Killing tensor is a gradient,
a Killing tensor (in general irreducible) can then be constructed.
In this paper it is shown that  
the severe restriction of orthogonality is unnecessary and thus it is
possible that many more Killing tensors can be constructed in this
way. We also extend, and in one case correct, some results
on Killing tensors constructed from a single conformal Killing
vector. Weir's result that, for flat space, there are 84 independent
conformal Killing tensors, all of which are reducible, is extended to
conformally flat spacetimes. In conformally flat spacetimes it is thus
possible to construct all the conformal Killing tensors and in
particular all the Killing tensors (which in general will not be
reducible) from conformal Killing vectors.
\end{resum}

\medskip
\section{Introduction}
In an $n$-dimensional Riemannian manifold a {\it Killing tensor} (of
order 2) is a symmetric tensor $K_{ab}$ satisfying
\begin{equation}
K_{(ab; c)} = 0
 \label{eq.1}
\end{equation}
A {\it conformal Killing tensor} (of order two) is a symmetric tensor 
$Q_{ab}$ satisfying
\begin{equation}
Q_{(ab; c)} = q_{(a}g_{bc)} \qquad {\rm with} \qquad
q_a = (Q_{,a}+ 2 Q^d_{\ a;d})/(n+2)
 \label{eq.2}
\end{equation}
where $Q=Q^d_{\ d}$.  In this paper only Killing and conformal Killing
tensors of order two will be considered so in future this
qualification will be assumed tacitly. 
When the associated {\it conformal vector}
$q_a \ne 0$, the conformal Killing tensor will be called {\it proper}
and otherwise it is a (ordinary) Killing tensor.  If $q_a$ is a
Killing vector, $Q_{ab}$ is referred to as a {\it homothetic
Killing tensor} (see \cite{Pr} for a discussion of such tensors). 
If the associated conformal vector $q_a = q_{,a}$ is 
the gradient of some scalar field $q$, then $Q_{ab}$ is called a 
{\it gradient conformal Killing tensor}.
It is easy to see that for each gradient conformal Killing tensor
$Q_{ab}$ there is an associated Killing tensor $K_{ab}$ given by
\begin{equation}
K_{ab} = Q_{ab} - qg_{ab}
 \label{eq.3}
\end{equation}
Such a Killing tensor is, of course, only defined up to the addition
of a constant multiple of the metric tensor.

Some authors (for example \cite{Kou, AM}) define a conformal Killing
tensor as a {\it trace-free} tensor $P_{ab}$ satisfying 
$P_{(ab; c)} = p_{(a}g_{bc)}$.
In fact there is no real contradiction or ambiguity between the two
definitions. If $P_{ab}$ is a trace-free conformal Killing tensor then
for any scalar field $\lambda$, $P_{ab} + \lambda g_{ab}$ is a
conformal Killing tensor and conversely if $Q_{ab}$ is a conformal
Killing tensor, its trace-free part $Q_{ab} - {1 \over n}Q g_{ab}$
is a trace-free Killing tensor.
In this paper the more general definition (\ref{eq.2}) will be
preferred and an explicit trace-free qualification will be added when
it is important to distinguish between the two definitions.

Killing tensors are of importance owing to their connection with
quadratic first integrals of the geodesic equations: if $p^a$ is
tangent to an affinely parameterised geodesic
(i.e. $p^a_{\ ;b}p^b=0$) it is easy to see that $K_{ab}p^ap^b$ is
constant along the geodesic. For {\it conformal} Killing tensors
$Q_{ab}p^ap^b$ is constant along {\it null} geodesics and here, of
course, only the trace-free part of $Q_{ab}$ contributes to the
constants of motion.
Both Killing tensors and conformal Killing tensors are also of
importance in connection with the separability of the Hamilton-Jacobi
equations and other partial differential equations. Separability will
not be considered further here; instead the reader is referred to the
literature on this subject (e.g.~\cite{Car, Ben} and the review
article \cite{BF}). 

A Killing tensor is said to be {\it reducible} if it can be written as
a constant linear combination of the metric and symmetrised products
of Killing vectors, i.e.
\begin{equation}
K_{ab}  =  a_0 g_{ab} +\sum^N_{I=1}\sum^N_{J=I} a_{IJ} 
\xi_{I(a}\xi_{|J|b)}
 \label{eq.5}
\end{equation}
where $\boldxi_I$ for $I = 1 \ldots N$ are the Killing vectors
admitted by the manifold and $a_0$ and $a_{IJ}$ for 
$1 \le I \le J \le N$  are constants.  Generally one is interested
only in Killing tensors which are not reducible since the quadratic
constant of motion associated with a reducible Killing tensor is
simply a constant linear combination of $p^ap_a$ and of pairwise
products of the linear constants of motion $\xi_{Ia}p^a$.

\section{Reducible Conformal Killing Tensors}
If $\boldchi_1$ and $\boldchi_2$ are independent conformal Killing
vectors with conformal factors $\vartheta_1$ and
$\vartheta_2$ (i.e. $\chi_{1(a;b)} = \vartheta_1 g_{ab}$ and 
$\chi_{2(a;b)} = \vartheta_2 g_{ab}$), it is easy to show
\begin{equation}
\chi_{1a}\chi_{1b} \qquad {\rm and} \qquad \chi_{1(a}\chi_{|2|b)}
 \label{eq.6}
\end{equation}
are conformal Killing tensors with associated conformal vectors
\begin{equation}
q_a= \vartheta_1\chi_{1a} \qquad {\rm and} \qquad 
q_a = (\vartheta_2\chi_{1a}+\vartheta_1\chi_{2a})/2
 \label{eq.7}
\end{equation}
Clearly the trace-free parts of these two tensors, namely 
$\chi_{1a}\chi_{1b}-{1 \over n}\chi_1^c\chi_{1c}g_{ab}$ and
$\chi_{1(a}\chi_{|2|b)} -{1 \over n}\chi_1^c\chi_{2c}g_{ab}$
are trace-free conformal Killing tensors. Note that if $\boldchi_1$ is
a proper conformal Killing vector, 
both the conformal Killing tensors in (\ref{eq.6}) are proper.  Note
also that if $\boldchi_1$ is a homothetic Killing vector with constant
conformal factor $\vartheta_1$ and $\boldchi_2$ is a Killing vector,
the second conformal Killing tensor in (\ref{eq.6}) is homothetic.

Koutras \cite{Kou} proved that if $\boldchi_1$ and $\boldchi_2$ were 
{\it orthogonal} conformal Killing vectors, then 
$Q_{ab}= \chi_{1a}\chi_{2b}+\chi_{1b}\chi_{2a}$
was a {\it trace-free} conformal Killing tensor and that, if
$\boldchi$ was a {\it null} conformal Killing vector, then
$Q_{ab} = \chi_a\chi_b$ was a {\it trace-free} conformal Killing
tensor. However, from the above considerations it is clear that the
assumptions of orthogonality or nullness are unnecessary; one can
simply take the symmetrised product of {\it any two} conformal Killing
vectors to obtain a conformal Killing tensor and then, if a trace-free
conformal Killing tensor is required, take the trace-free part of this
tensor. 

It is easy to see that any scalar multiple of the metric
tensor $\lambda g_{ab}$ is a conformal Killing tensor with associated
conformal vector $q_a = \lambda_{,a}$.  A conformal Killing tensor
$Q_{ab}$ is said to be {\it reducible} if it can be written as a
linear combination of the metric and symmetrised products of conformal 
Killing vectors:
\begin{equation}
Q_{ab}  =  \lambda g_{ab} +\sum^N_{I=1}\sum^M_{J=I} a_{IJ} 
\chi_{I(a}\chi_{|J|b)}
 \label{eq.9}
\end{equation}
where $\boldchi_I$, for $I = 1 \ldots M$, are the conformal Killing vectors
admitted by the manifold and $a_{IJ}$ for $1 \le I \le J \le M$  are
constants. Note that here, unlike in the definition of
reducibility in (\ref{eq.5}), the coefficient 
$\lambda$ multiplying the metric is not constant. The conformal
vector $q_a$ associated with the conformal Killing tensor (\ref{eq.9})
is
\begin{equation}
q_a = \lambda_{,a} + \sum^M_{I=1}\sum^M_{J=I}
{1 \over 2}a_{IJ}(\vartheta_I \chi_{Ja}+\vartheta_J \chi_{Ia})
  \label{eq.10}
\end{equation}
Thus, if the conformal Killing vectors of a manifold are known, the
following prescription for finding Killing tensors suggests itself:
construct all the reducible conformal Killing tensors as in
(\ref{eq.9}) and then determine which of these (if any) are gradient
(by finding the linear subspace for which $q_{[a,b]}= 0$) and, for this
subspace, construct the associated Killing tensors as in (\ref{eq.3}).
Note that if $Q_{ab}$ is a gradient conformal Killing tensor, then so
is $Q_{ab} +\mu g_{ab}$ for any scalar field $\mu$. Thus, without loss of
generality in this construction, $\lambda$ in (\ref{eq.9}) may be
assumed to be zero.

If the metric admits $N$ independent Killing vectors
$\boldxi_1, \ldots, \boldxi_N$, the linear space of all reducible
conformal Killing tensors given by (\ref{eq.9}) contains a linear subspace of
reducible Killing tensors of the form (\ref{eq.5}).
Since reducible Killing tensors are of little interest, these
reducible tensors can be excluded if a basis of the 
conformal Killing vectors is chosen to be of the form 
$\boldxi_1, \ldots, \boldxi_N, \boldchi_{N+1}, \ldots, \boldchi_M$
and then only reducible conformal Killing tensors of the following
form are considered:
\begin{equation}
Q_{ab} =  \sum^N_{I=1}\sum^M_{J=N+1} a_{IJ}
    \xi_{I(a} \chi_{|J|b)} + \sum^M_{I=N+1}\sum^M_{J=I} a_{IJ}
    \chi_{I(a} \chi_{|J|b)}
 \label{eq.11}
\end{equation}
Occasionally the Killing tensors
constructed in this way will be reducible and in general it is
necessary to check whether they can be expressed in the form
(\ref{eq.5}).

\section{Killing Tensors from 1 Conformal Killing Vector}

\noindent {\bf Theorem 1.}
Any manifold which admits a proper non-null conformal Killing vector
field $\boldchi$ which is geodesic (that is 
$\chi_{a;b}\chi^b = \lambda\chi_a$) also admits the Killing tensor
$K_{ab}=\chi_a\chi_b-{1 \over 2}\chi^2g_{ab}$, where $\chi^2 = \chi^c\chi_c$.

\noindent {\bf Proof.} Contracting the equation
$\chi_{(a;b)}=\vartheta g_{ab}$ with $\chi^a\chi^b$ gives
$\lambda\chi^2 = \vartheta\chi^2$. Hence as $\boldchi$ is non-null,
$\lambda=\vartheta$. Contracting $\chi_{(a;b)}=\vartheta g_{ab}$ with
$\chi^b$ gives $\vartheta \chi_a = {1 \over 2}(\chi^2)_{,a} $ and
hence from equations (\ref{eq.3}, \ref{eq.6} \& \ref{eq.7}),
$K_{ab}=\chi_a\chi_b -{1 \over 2}\chi^2g_{ab}$ is a Killing tensor.

Koutras \cite{Kou} proved this result for {\it homothetic} Killing
vectors only. Our proof is more general and direct
as it does not rely on the introduction of a particular co-ordinate
system.
Koutras claimed that the result was valid for null $\boldchi$, but
this is false as the following counter-example shows.  Consider the metric:
\begin{equation}
ds^2 = e^{2u}(2A(x,y,v)dudv +dx^2+dy^2)
 \label{eq.12}
\end{equation}
It is easy to see that $\chi^a$ is a null homothetic Killing vector
with conformal factor $\vartheta =1$.  As $\chi^a$ is a null conformal
Killing vector, it is necessarily geodesic.  The associated conformal
Killing tensor $Q_{ab}$ and conformal vector $q_a$ are 
\begin{equation}
Q_{ab} =  A^2e^{4u}\delta^v_a\delta^v_b \qquad q_a= e^{2u}\delta^v_a
 \label{eq.13}
\end{equation}
A straightforward calculation shows that $q_a$ is not a gradient.

The following theorem generalises that of Amery \& Maharaj
\cite{AM} for homothetic Killing vectors.

\noindent{\bf Theorem 2.}
Any manifold which admits a conformal Killing vector $\boldchi$ that
is a gradient vector, also admits the Killing tensor
$K_{ab}=\chi_a\chi_b-{1 \over 2}\chi^2g_{ab}$.

\noindent{\bf Proof.} As $\chi_a$ is a gradient, $\chi_{[a;b]} =0$ and
thus $\chi_{a;b} = \vartheta g_{ab}$ and contraction with $\chi^b$
gives $\vartheta \chi_b ={1 \over 2}(\chi^2)_{,b}$ and the result
follows as in the proof of theorem 1.

For non-null vectors theorem 2 also follows from theorem 1
since gradient conformal Killing vectors are geodesic. For the null
case a gradient conformal Killing vector is
a Killing vector and so the associated Killing tensor is necessarily
reducible.

\section{Conformal Transformations}
If $\chi^a$ is a conformal Killing vector of the metric $g_{ab}$ with
conformal factor $\vartheta$, it is also a conformal Killing vector of
the metric $\tilde g_{ab} = e^{2\Omega} g_{ab}$ with
conformal factor $\tilde \vartheta = \vartheta + \Omega_{,c}\chi^c$.
The analogous result for conformal Killing tensors is:

\noindent {\bf Theorem 3.} If $Q^{ab}$ is a conformal Killing tensor
satisfying  $\nabla^{(a} Q^{bc)}=q^{(a}g^{bc)}$, then $Q^{ab}$ is also
a conformal Killing tensor of the conformally related metric  
$\tilde g_{ab} =  e^{2\Omega}  g_{ab}$.  $Q^{ab}$ satisfies 
$\tilde  \nabla^{(a} Q^{bc)}=\tilde q^{(a} \tilde g^{bc)}$,
where $\tilde q^a = q^a  +2 \Omega_{,d}Q^{da}$.

\noindent {\bf Proof.} The proof is straightforward involving an evaluation of
$\tilde  \nabla^{(a} Q^{bc)}$ using
\begin{equation}
\tilde \Gamma^a_{bc} = \Gamma^a_{bc} + \delta^a_b \Omega_{,c}
             + \delta^a_c \Omega_{,b} -\Omega^{,a}g_{bc}
 \label{eq.14}
\end{equation}

From this theorem and the corresponding theorem for conformal Killing
vectors it follows that the number of linearly independent {\it
trace-free} conformal Killing tensors is invariant under conformal
change of metric and the number of linearly independent {\it
reducible} trace-free conformal Killing tensors is similarly invariant.
Note the trace-free qualification here; owing to the freedom to add
arbitrary multiples of the metric, the number of independent
conformal Killing tensors in a manifold is infinite. 

The maximal number of independent trace-free
conformal Killing tensors admitted by an $n$-dimensional Riemannian
manifold is ${1 \over 12}(n-1)(n+2)(n+3)(n+4)$ and this limit is
achieved when the metric is flat \cite{W}. In this case all the conformal
Killing tensors are reducible.  Noting theorem 3, we see that
these results are valid for any conformally flat metric. Thus, in
particular, all Killing tensors in a conformally flat spacetime may be
constructed from products of conformal Killing vectors.

Amery \& Maharaj \cite{AM} used Koutras' methods to find Killing
tensors in Robertson-Walker metrics (which are conformally
flat). They were  able to construct only 39 of the 84 possible
reducible conformal Killing tensors as they used only mutually 
orthogonal conformal Killing vectors in their construction. Thus their
results are likely to be incomplete. Currently work is in progress
investigating Killing tensors in conformally flat spacetimes and
the results will be presented elsewhere.

\end{document}